\documentclass[11pt]{article}
\usepackage[utf8]{inputenc}
\usepackage{amsmath, amssymb}
\usepackage{graphicx}
\usepackage{cite} 
\usepackage{hyperref}
\title{Super-resolution AWGs based on the Moir\'e  effect}
\author{Gabriella Cincotti\\
Department of Civil, Computer and Aeronautical Engineering \\
University Roma Tre, via  Vito Volterra 62,  I-00146 Rome Italy\\
gabriella.cincotti@uniroma3.it} 

\date{\today} 
\begin{document}
\maketitle

\begin{abstract} 
Enhancing spectral resolution in a conventional arrayed waveguide grating AWG requires a longer differential optical path, which expands the device footprint and increases sensitivity to phase noise. To overcome this limitation, we propose an innovative approach that maintains the AWG layout and  introduces an additional splitter  at  the device input. Leveraging the Moir\'e effect, generated by two gratings with slightly different pitches, we demonstrate that the spacing of $7$ channels multiplexer/demultiplexer can be reduced to $1$ GHz at the central wavelength of $1550$ nm. We present the theoretical foundation for a super-resolution AWG and provide the corresponding design guidelines.
\end{abstract}

\section{Introduction}
Arrayed Waveguide Gratings (AWGs) are fundamental optical devices in wavelength division multiplexing (WDM), which is the dominant transmission approach in optical networks due to its numerous advantages, including high capacity, long transmission distances and reliability \cite{Mukherje}. AWGs serve as key planar lightwave circuit (PLC)-based components in  WDM systems, and they are  used for optical multiplexing/demultiplexing  (MUX/DEMUX),  filtering, wavelength metrology, and other signal processing applications \cite{tu, chen, li, kataoka, cincotti1, cincotti2}.
Beyond optical communications, AWGs have found applications in various other optical fields, including spectroscopy \cite{murakowski} and astronomy \cite{gatkine}. They are employed in integrated photonic spectrographs for space telescopes, where minimizing mass and volume is critical \cite{stoll,stoll2}. Additionally, AWGs are increasingly utilized in biophotonics, such as for near-infrared spectroscopy \cite{kawagoe} and optical coherence tomography (OCT) \cite{rank,seyringer}.

The concept of AWG  was introduced by M. Smit in 1988 \cite{smit}  and by C. Dragone in 1991, \cite{dragone} and it is based on the phased-array principle. An array of $M+1$ waveguides, each with a progressively increasing optical path induces phase shifts that separate incoming wavelengths across $N$ output channels. The two key  parameters are spectral resolution, \textit{i.e.} the ability  to resolve closely spaced spectral lines, and the wavelength range. 
The major  challenge in AWG design is achieving high spectral resolution  while ensuring a wide operational bandwidth. This trade-off arises because increasing the resolution requires a larger optical path difference $\Delta L$, which increases the device footprint and introduces fabrication complexities and phase errors. 

Miniaturized integrated spectrometers capable of resolving optical spectra with high resolution are highly promising for widespread applications in fields such as medical diagnostics \cite{ferrari}, environmental monitoring, and hazard detection \cite{witinski}. These  devices are essential for advanced techniques like OCT and Brillouin spectroscopy \cite{Rioboo}. However, achieving spectral resolutions of $1$  GHz or better remains challenging in current state-of-the-art AWGs due to fabrication constraints. These include the need for long optical path lengths, increased device footprint, and high sensitivity to environmental factors, especially temperature variations.

To address these inherent constraints, various different techniques have been proposed \cite{van2023custom}. One approach involves cascading two AWGs with slightly different free spectral ranges (FSRs), leveraging the Vernier effect to achieve finer wavelength discrimination \cite{takata}. While the cascading configuration enables flexible WDM devices with a high channel count $N$ and narrow channel spacing, the device size remains an issue. This can be mitigated by interleaving, exploiting the periodic property of an AWG: a high-resolution device serves as the first stage, while several coarse filters are used in the second stage \cite{akca}.

An other approach integrates external optical elements, such as bulk optics or fiber-based filters, to further enhance resolution \cite{akca2}. In this case, the AWG output waveguides  are removed and the output spectrum is detected by a pixel-based imaging system. The overall spectral resolution in this configuration depends on the combined AWG resolution and detector pixel size. Additionally, computational algorithms have been proposed to surpass the diffraction-limited resolution of conventional AWGs \cite{gao}. Despite these innovations, achieving a spectral resolution finer than 1 GHz at the telecom wavelength $1550$ nm remains a significant challenge \cite{zheng, hibino, gehl}.

In this paper, we propose a novel AWG architecture to enhance spectral resolution, based on the Moir\'e  effect, which arises when two optical gratings with slightly different periods are used. This approach has been successfully implemented in a bulk-optic spectrometer, demonstrating its potential to surpass conventional resolution limits \cite{Konishi}. Here, we adapt the same principle to an AWG-based architecture, introducing a novel design that exploits Moir\'e  interference patterns to achieve high-resolution resolution. We theoretically demonstrate that this approach enables $1$ GHz spectral resolution, opening a new pathway toward ultra high-resolution integrated photonic spectrometers.

 In the proposed device, an additional Y-branch splitter or a slab coupler is integrated into a conventional AWG configuration.  Although this increases the overall footprint, the device remains more compact than the cascaded or interleaved AWG architectures.

The paper is organized as follows: In Sec. \ref{sec2}, we describe the basic principles of a conventional spectrometer and the super-resolution spectrometer proposed by T. Konishi \cite{Konishi}. In Sec. \ref{sec3}, we apply the Moir\'e  effect to a custom AWG, adding an input waveguide grating, and evaluate the corresponding spectral resolution. An accurate AWG model, including diffraction effects, is presented in Sec. \ref{sec4}, along with detailed design guidelines. Finally, conclusions are provided in Sec. \ref{sec5}.

\section{Super-resolution spectrometer}
\label{sec2}
\begin{figure}[!t]
\centering
\includegraphics[width= \textwidth]{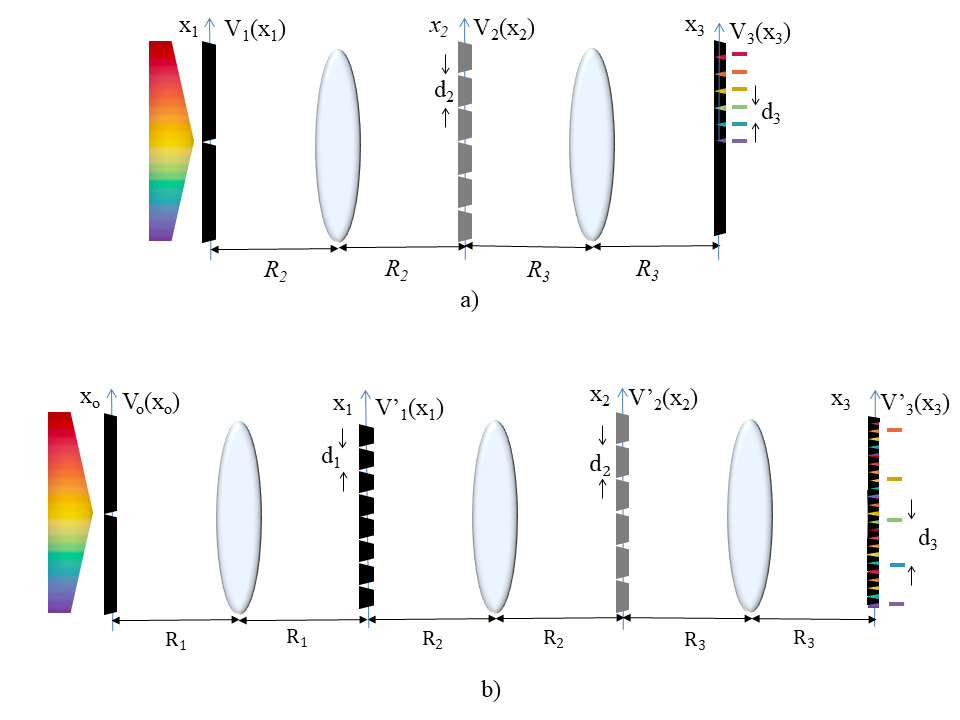}
\caption{a) Conventional spectrometer. b) Super-resolution spectrometer.}
\label{fig1}
\end{figure}

We start illustrating the basic principle of a conventional spectrometer reported in Fig.~\ref{fig1}a, to better clarify  the innovation of T. Konishi's approach. 
A polychromatic plane wave, with constant amplitude $A$ and central wavelength $\lambda_c$, illuminates the entrance slit, that has a transmittance profile $b_1(x_1)$. A dispersive grating with pitch $d_2$ is positioned at the focal plane between two lenses of focal lengths $R_2$ and $R_3$. The input field distribution $V_1(x_1) = A b_1(x_1)$ is Fourier transformed by the first lens at the spatial frequency $x_2/\lambda R_2$ and is transmitted by the diffraction grating with transmittance
\begin{equation}
\label{taugrating}
    \tau_2(x_2) = e^{i2\pi \bar{m} \frac{ x_2}{d_2}}.
\end{equation}

Here $i$ is the imaginary constant and $\bar{m}$ the diffraction order. The second lens performs another Fourier transform at the spatial frequency $x_3/\lambda R_3$, and the output field is proportional to
\begin{equation}
\label{V3gratingnormale}
    V_3(x_3) = A b_1\left(\frac{x_3 R_2}{R_3}-\frac{ \bar{m}\lambda R_2}{d_2}\right).
\end{equation}

The input light field is spatially dispersed at the output plane, according to wavelength and an array of $N$ photodetectors, spaced $d_3$ apart, measures the spectral components. The spectral resolution
\begin{equation}
\label{spectralresolution}
    \Delta \lambda = \frac{d_3 d_2}{\bar{m} R_3}.
\end{equation}
can be improved with higher $\bar{m} $, narrower entrance slit, and smaller values of $d_3$ and $d_2$. Increasing the diffraction order $\bar{m}$ is quite difficult due to several fundamental constraints related to angular limitation, efficiency, and fabrication limits. In addition, reducing the photodetector pixel and the entrance slit is challenging and affects the spectrometer sensitivity.

To improve the spectral resolution, T. Konishi proposed and fabricated an innovative spectrometer leveraging the Moir\'e effect, achieving a tenfold increase in spectral resolution to $0.31$ nm at a center wavelength $\lambda_c=1550$ nm \cite{Konishi}. 
The schematic of this super-resolution spectrometer is illustrated in Fig.~\ref{fig1}b, with detailed descriptions available in \cite{Konishi}. In this new design, the entrance slit is replaced by an array of $N$ slits with pitch $d_1$, resulting in the input field distribution
\begin{equation}
    V'_1 (x_1) = A\sum_{n=-(N-1)/2}^{(N-1)/2} b_1 (x_1 - n d_1).
\end{equation}
The output field  becomes proportional to
\begin{equation}
\label{V3supergrating}
    V'_3 (x_3) = A\sum_{n=-(N-1)/2}^{(N-1)/2} b_1 \left(\frac{x_3 R_2}{R_3} - \frac{\bar{m} \lambda R_2}{d_2} - n d_1 \right),
\end{equation}
so that this configuration spatially disperses $N$ copies of the entrance slits, according to wavelength. By spacing the photodetectors of 
$d_3$,
the spectral resolution is enhanced as
\begin{equation}
\label{superesolution}
    \Delta \lambda' = \left(\frac{d_3}{R_3}-\frac{d_1}{R_2}\right)\frac{d_2}{\bar{m}} = \Delta \lambda \left( 1 - \frac{R_3 d_1}{R_2 d_3} \right)=  \frac{\Delta \lambda}{F}
\end{equation}
where
\begin{equation}
\label{d3}
F=\frac{R_2d_3} {R_2d_3-R_3d_1}>1,
\end{equation}
 is a magnification parameter.
Therefore, a suitable selection of the pitches $d_3$ and $d_1$, along with the focal lengths $R_3$ and $R_2$ can significantly enhance the spectral resolution. It is also important to remark that the additional input grating produces $N$ replica of the  slit profile $b_1\left(x_3  \right)$  at the output plane,  spaced by $x_3= R_3 d_1/R_2$. Therefore the super-resolution spectrometer presents the  
\begin{equation}
\label{FSR'}
    FSR' = \frac{d_1 d_2}{\bar{m}R_2},
\end{equation}
and to ensure that all output frequency channels fit within 
FSR' (\textit{i.e.} $ N \Delta \lambda ' \le$  FSR'), it is required that  $F \ge N+1$. 
\section{Super-resolution AWG}
\label{sec3}
\begin{figure}[!t]
\centering
\includegraphics[width= \textwidth]{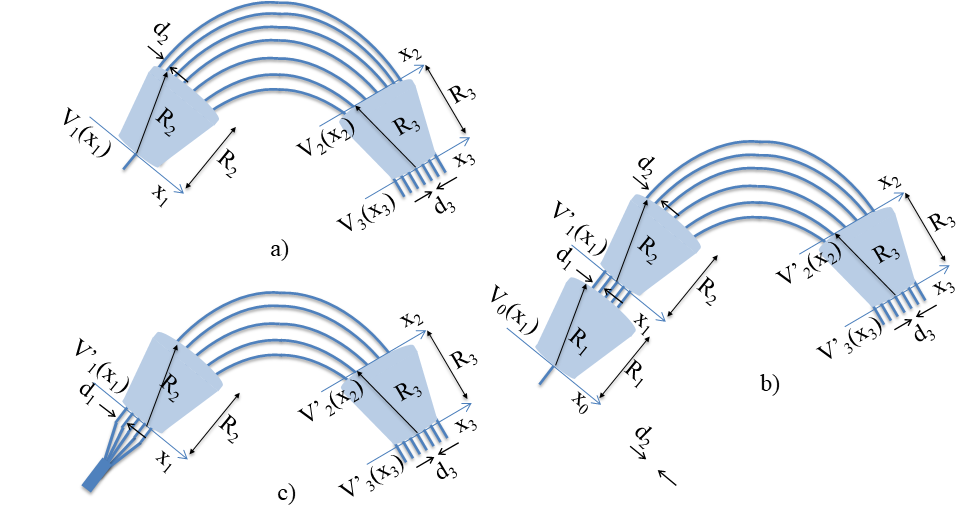}
\caption{a) Conventional AWG. b) Super-resolution AWG with an additional slab coupler. c) Super-resolution AWG with an input  Y-branch.}\label{fig2}
\end{figure}
While conventional AWGs  are generally designed symmetrically, with $N$ input and output ports, for clarity, we initially refer to a configuration with a single input port and $N$ output ports, as depicted in Fig. \ref{fig2}a. A conventional AWG layout comprises an array of $M+1$ waveguides with an incremental length $\Delta L$ and pitch $d_2$. The two slab couplers, which behave as lenses with focal length $R_1$ and $R_2$, perform the Fourier transform of the light field $V_1(x_1) =A b_1(x_1)$ at the input waveguide. Here,  $b_1(x_1)$ represents the  mode profile of the input waveguide, and $A$ is a suitable constant. The transmittance of the arrayed grating is 
\begin{eqnarray}
\label{tauawg}
    \tau'_2(x_2)= \sum_{m=-M/2}^{M/2} \delta (x_2 - m d_2) e^{i 2\pi m \Delta L/\lambda} \nonumber\\
    =\sum_{m=-M/2}^{M/2} \delta (x_2 - m d_2)\cdot \tau_2(x_2).
\end{eqnarray}
where $\delta(x_2)$ is the Dirac delta, $\bar{m}=\frac{\Delta L}{\lambda}$ and  $\lambda$ is the wavelength in the dielectric. A more accurate model, that takes the arrayed waveguide  mode profile $b_2(x_2) $ into account, is presented in the following paragraph. 

The output field is proportional to
\begin{eqnarray}
\label{V3AWGnormale}
V_3(x_3) = A b_1\left(\frac{x_3 R_2}{R_3}\right)  * \sum_{m=-M/2}^{M/2} e^{-i \frac{2\pi m d_2 }{\lambda R_3}\left(x_3 - \frac{ \bar{m} \lambda R_3}{d_2}\right)}\nonumber\\
 =A b_1\left(\frac{x_3 R_2}{R_3}\right)   * \frac{\sin\left[ \left(M+1\right)\frac{\pi d_2 }{\lambda R_3}\left(x_3 - \frac{ \bar{m} \lambda R_3}{d_2}\right)\right]}{\sin\left[ \frac{\pi d_2 }{\lambda R_3}\left(x_3 - \frac{ \bar{m} \lambda R_3}{d_2}\right)\right]},
\end{eqnarray}
where $*$ denotes convolution. It is well known that the conventional AWG configuration produces replicas spaced  by $\lambda R_3/d_2$ at the output plane, and the corresponding  FSR is determined by setting
 \begin{eqnarray}
 \frac{\bar{m}\lambda'R_3}{d_2}=\frac{\left(\bar{m}+1\right)\lambda R_3}{d_2},
   \end{eqnarray}
that is 
 \begin{eqnarray}
 \label{FSR}
 FSR=\lambda' -\lambda=\frac{\lambda}{\bar{m}}.
   \end{eqnarray}

If the waveguides are spaced by $d_3$  at the output plane, the spectral resolution is given by Eq. (\ref{spectralresolution}). 
 
The number of arrayed waveguides $M$ and their spacing $d_2$  are critical design parameters that influence both crosstalk and insertion losses. The spacing $d_2$ must be large enough to prevent evanescent coupling between adjacent waveguides; however, if it is too large, optical power can be lost, leading to increased insertion loss. A smaller value of $d_2$ helps reduce crosstalk and improve spectral performance. Increasing the number of arrayed waveguides $M$ enhances the filtering function and  reduces crosstalk. However, it also increases the overall device size and contributes to higher insertion losses.
 
 To integrate the Moir\'e effect into the design of the AWG, we replace the single input waveguide of Fig.~\ref{fig2}a with a grating  consisting of $N$ waveguides, with mode profile $b_1(x_1)$ and separated by a distance $d_1$, as illustrated in Fig.~\ref{fig2}b. It is important to remark that in a conventional AWG with $N$ input ports, the field distribution at the input ports varies because it corresponds to frequency channels that are multiplexed. In our case, the $N$ input waveguides have the same field distribution and, to ensure their uniform excitation,  we introduce either an additional confocal slab coupler with focal length $R_1$ (see Fig.~\ref{fig2}b) or a $1\times N$ splitter (see Fig.~\ref{fig2}c).  In the latter case, all $N$ input waveguides share the same field distribution and
 \begin{equation}
V'_1 (x_1) = A  \sum_{n=-(N-1)/2}^{(N-1)/2} b_1 (x_1 - n d_1),
\end{equation}
where $A$ is a suitable constant.
In the following paragraph, we present a more comprehensive analysis, that also accounts for diffraction effects introduced by the additional input slab coupler. 

 The light field  transmitted by the arrayed grating is proportional to
\begin{eqnarray}
\label{V2primo}
&&V'_2 (x_2) =A  \tilde{b}_1 \left( \frac{x_2 }{\lambda R_2} \right)\nonumber \\
&&\cdot \sum_{n=-(N-1)/2}^{(N-1)/2} \sum_{m=-M/2}^{M/2} \delta\left(x_2 -m d_2\right) e^{i 2\pi  x_2 \left(\frac{n d_1}{\lambda R_2} +\frac{\bar{m}}{d_2}\right)}\nonumber \\
\end{eqnarray}
where $\tilde{b}_1 \left( \frac{x_2 }{\lambda R_2}\right)$ is the Fourier transform of the waveguide mode profile $b_1(x_1)$. 
Finally, we can evaluate the output field, that is proportional to
\begin{eqnarray}
\label{uscita}
&&V'_3(x_3) = A  b_1\left(\frac{x_3 R_2}{R_3}\right) \nonumber\\
&&*  \sum_{n=-(N-1)/2}^{(N-1)/2}\sum_{m=-M/2}^{M/2} e^{-i \frac{2\pi m d_2 }{\lambda R_3}\left(x_3 - \frac{n d_1 R_3}{  R_2}-\frac{ \bar{m} \lambda R_3}{d_2}\right)}\nonumber\\
&& =A b_1\left(\frac{x_3 R_2}{R_3}\right) \nonumber\\
&& *  \sum_{n=-(N-1)/2}^{(N-1)/2}\frac{\sin\left[ \left(M+1\right)\frac{\pi d_2 }{\lambda R_3}\left(x_3 - \frac{n d_1 R_3}{  R_2}-\frac{ \bar{m} \lambda R_3}{d_2}\right)\right]}{\sin\left[ \frac{\pi d_2 }{\lambda R_3}\left(x_3 - \frac{n d_1 R_3}{  R_2}-\frac{ \bar{m} \lambda R_3}{d_2}\right)\right]}.\nonumber\\
\end{eqnarray}

By spacing the output waveguides of $d_3$ given in Eq. (\ref{d3}), the spectral resolution is enhanced as per Eq.~(\ref{superesolution}). 
The FSR remains unchanged and is given by  Eq. (\ref{FSR}), but the super-resolution AWG presents the additional FSR'$<$FSR due to the Moir\'e effect given in Eq. (\ref{FSR'})

\begin{equation}
    FSR' = \frac{d_1 d_2}{\bar{m}R_2}< FSR=\frac{\lambda}{\bar{m}}.
\end{equation}

To illustrate the Moir\'e effect in an AWG architecture, we consider a case that is not commonly used in MUX/DEMUX design, but provides a simple and insightful example. The AWG design guidelines for a super-resolution MUX/DEMUX are provided in the next paragraph. In this example, the Moiré replicas are uniformly spaced within the FSR \textit{(i.e.} FSR$=N \cdot$ FSR') and we set
\begin{equation}
\label{d1}
d_1 = \frac{\lambda R_2 }{d_2 N}.
\end{equation}

Additionally, setting $F=N+1$ in Eq. (\ref{d3}), we place the output waveguide spaced of
\begin{equation}
\label{dd3}
d_3 = \frac{R_3 d_1}{R_2} \frac{F}{F-1} = \frac{R_3 d_1}{R_2} \left( 1 + \frac{1}{N} \right), 
\end{equation}
and the spectral resolution of Eq.~(\ref{superesolution}) becomes
\begin{equation}
\label{dek}
\Delta \lambda' = \frac{\Delta \lambda}{N+1}.
\end{equation}

Finally, substituting Eqs. (\ref{d1}) and (\ref{dd3}), into Eq. (\ref{dek}), we have
\begin{equation}
\Delta \lambda' = \frac{\lambda}{\bar{m} N^2} = \frac{\lambda^2}{\Delta L N^2},
\end{equation}
which represents an enhancement by a factor of $N$ compared to a conventional AWG with the same incremental path length $\Delta L$. 

Substituting these parameters into Eq.~(\ref{uscita}), the final expression for the output field is obtained
\begin{eqnarray}
\label{uscita2}
&&V'_3(x_3) = Ab_1\left(\frac{x_3 R_2}{R_3}\right) \nonumber\\
&& *  \sum_{n=-(N-1)/2}^{(N-1)/2}\frac{\sin\left[\pi \left(M+1\right)\left(\frac{ x_3d_2 }{\lambda R_3} - \frac{n }{N }-\bar{m}\right)\right]}{\sin\left[\pi \left(\frac{ x_3d_2 }{\lambda R_3} - \frac{n }{N  }-\bar{m}\right)\right]}.
\end{eqnarray}

\begin{figure}[!t]
\centering
\includegraphics[width= \textwidth]{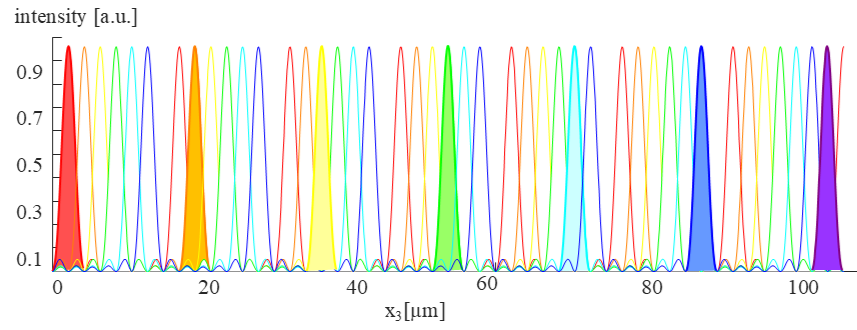}
\caption{a) Field distribution $V'_3(x_3)$ at the output plane.  The frequency channels, spaced of $1$ GHz and transmitted at the $N=7$ output ports are highlighted with a filled color under the curve.}
\label{fig3}
\end{figure}

The output field distribution given by Eq. (\ref{uscita2}) is plotted in Fig. ~\ref{fig3} setting the AWG parameters as $N=7$,  $M=50$, $\lambda_c=1550  \mu $m, $d_1 = 15   \mu $m, $d_2=20  \mu $m, $d_3 =  17.14 \mu $m, $R_2=R_3=1354.8 \mu $m,  $\Delta L=6118.2 \mu $m. For clarity, we have neglected the effects of the waveguide field profile, putting $ b_1 (x_3) = 1$. 

\section{Accurate model and design guidelines}
\label{sec4}
In a more accurate model, we refer to the layout of Fig.~\ref{fig2}b, where an additional slab coupler with radius $R_1$ is inserted, so that 
\begin{equation}
\label{V1primo}
V'_1 (x_1) = A \tilde{b}_0 \left( \frac{x_1}{\lambda R_1} \right) \cdot \sum_{n=-(N-1)/2}^{(N-1)/2} b_1 (x_1 - n d_1).
\end{equation}

$A \tilde{b}_0 (x_1 / \lambda R_1)$ is the Fourier transform of the field distribution $V_0 (x_0) = A b_0 (x_0)$, and $b_0(x_0)$ is the  mode profile.

For simplicity, we assume that the light coupled into each waveguide is proportional to the field distribution at its center.
Therefore, we rewrite Eq. (\ref{V1primo}) as
\begin{equation}
V'_1 (x_1) = A  \sum_{n=-(N-1)/2}^{(N-1)/2} \tilde{b}_0 \left( \frac{n d_1}{\lambda R_1} \right) b_1 (x_1 - n d_1),
\end{equation}
and the light transmitted by the arrayed grating becomes
\begin{eqnarray}
V'_2 (x_2) =A   \sum_{n=-(N-1)/2}^{(N-1)/2}\sum_{m=-M/2}^{M/2}\tilde{b}_0 \left( \frac{n d_1}{\lambda R_1} \right)\nonumber \\
\cdot \tilde{b}_1 \left( \frac{m d_2 }{\lambda R_2} \right) b_2\left(x_2 -m d_2\right) e^{-i 2\pi  x_2 \left(\frac{n d_1}{\lambda R_2} -\frac{\bar{m}}{d_2}\right)}.
\end{eqnarray}
Finally, the output field of Eq. (\ref{uscita2}) becomes
\begin{eqnarray}
\label{V3final2}
&&V'_3(x_3) = A\tilde{b}_2\left(\frac{x_3}{\lambda R_3}\right)\cdot\sum_{n=-(N-1)/2}^{(N-1)/2}\sum_{m=-M/2}^{M/2}\tilde{b}_0 \left( \frac{n d_1}{\lambda R_1} \right)\nonumber\\
&&\cdot \tilde{b}_1 \left( \frac{m d_2 }{\lambda R_2} \right) 
 e^{-i \frac{2\pi m d_2 }{\lambda R_3}\left(x_3 - \frac{n d_1 R_3}{  R_2}-\frac{ \bar{m} \lambda R_3}{d_2}\right)}.
 \end{eqnarray}

We assume that the waveguide mode profiles have Gaussian distributions
\begin{eqnarray}
b_0(x_0) = \frac{1}{\sqrt{\pi}w_0} e^{-x_0^2/w_0^2}\nonumber\\
b_1(x_1) = \frac{1}{\sqrt{\pi}w_1} e^{-x_1^2/w_1^2}\nonumber\\
b_2(x_2) = \frac{1}{\sqrt{\pi}w_2} e^{-x_2^2/w_2^2}
 \end{eqnarray}
where $w_0$, $w_1$ and $w_2$ are the spot sizes. Substituting into Eq. (\ref{V3final2}), we obtain
\begin{eqnarray}
\label{V3_1}
&&V'_3(x_3) = Ae^{-\left(\frac{\pi w_2 x_3}{\lambda R_3}\right)^2}\sum_{n=-(N-1)/2}^{(N-1)/2}e^{-\left(\frac{\pi w_0 n d_1}{\lambda R_1}\right)^2}\nonumber\\
&&\cdot \sum_{m=-M/2}^{M/2}e^{-\left(\frac{\pi w_1 m d_2}{\lambda R_2}\right)^2}
  e^{-i \frac{2\pi m d_2 }{\lambda R_3}\left(x_3 - \frac{n d_1 R_3}{  R_2}-\frac{ \bar{m} \lambda R_3}{d_2}\right)}.
 \end{eqnarray}
 
If an additional slab coupler is included in the AWG layout, it is crucial that its curvature radius $R_1$ is   sufficiently large to minimize diffraction effects. For the same reason, the spot-size  $w_0$ of the mode profile $b_0(x_0)$  should be chosen as small as possible. However, both of these conditions unfortunately reduce the input power coupled into the AWG. To mitigate these effects, the additional slab coupler could be replaced with a $1 \times N$ splitter, as shown in Fig. \ref{fig2}c.

 \begin{figure}[!t]
\centering
\includegraphics[width= \textwidth]{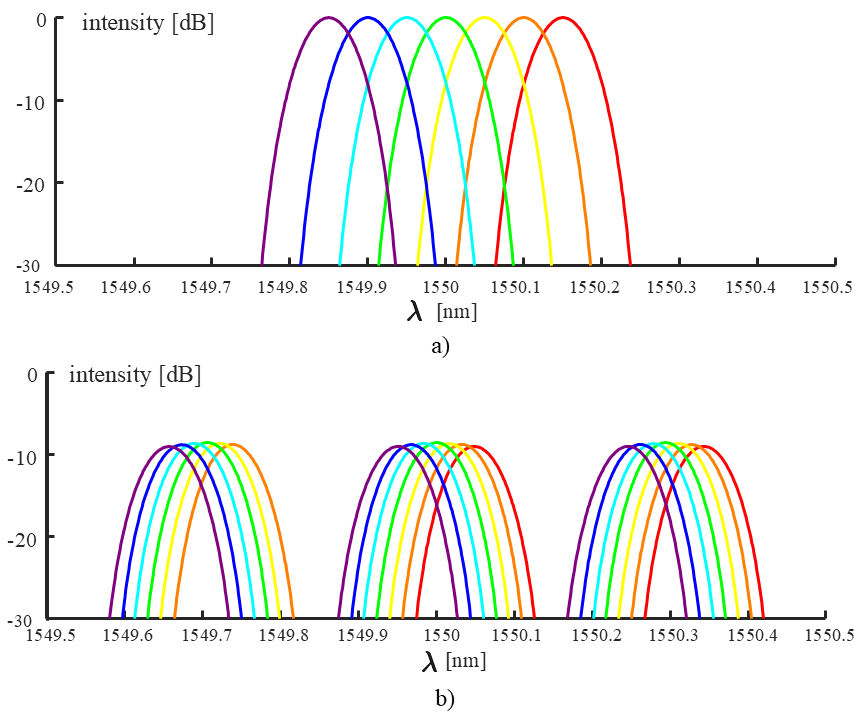}
\caption{a) Spectra transmitted at the $N=7$ output ports. a) Conventional AWG. b) Super-resolution AWG.}
\label{fig4}
\end{figure}

The design guidelines for a super-resolution AWG are similar to those of a custom AWG \cite{smit2}, with the addition of an input array designed with pitch $d_1$ to achieve the Moir\'e effect and the spectral resolution of Eq. (\ref{superesolution}).

The arrayed waveguide spacing $d_2$ should be chosen as small as possible, to reduce crosstalk. However a small value, there evanescent field coupling among the wavelength occurs that increases phase errors and the crosstalk. Likewise, all arrayed waveguides must be properly illuminated, so their number $M$ is selected accordingly, as described in \cite{seyringerbook}. 
\begin{equation}
M+1 > \frac{\lambda R_2}{w_1 d_2}. 
\end{equation}

Increasing $M$ reduces crosstalk but also increases the footprint. Additionally, a larger spot size $w_1$ impacts crosstalk by broadening the filter functions, thereby reducing channel isolation.

Loss non-uniformity, \textit{i.e.}, variations in transmitted power across the output ports, primarily arises from the mode profile $b_2(x_2)$. To mitigate this effect, reducing the waveguide width and the corresponding spot size $w_2$, along with increasing the curvature radius $R_3$, is beneficial. 

The design process begins with knowledge of the fabrication parameters, such as the PLC platform, waveguide width as well as waveguide effective refractive index $n_{eff}$, and slab refractive index $n_{s}$. Key design inputs include the number of output channels $N$, the central wavelength, and of course the required spectral resolution $\Delta \lambda'$. 

To provide a more insightful example, instead of designing a novel super-resolution device, we modify the layout of  the conventional AWG designed and fabricated in Ref. \cite{tu} and incorporate the input grating with pitch $d_1$ and therefore the Moiré effect. We consider the same parameters: central wavelength $\lambda_c=1550$ nm, waveguide effective refractive index $n_{eff}=2.279$,  slab refractive index $n_{s}=2.039$, number of arrayed waveguides $M=400$, incremental length $\Delta L=52.61 \mathrm{\mu m}$, waveguide spot sizes $w_1=w_2=5 \mathrm{\mu m}$, slab curvature radii $R_2=R_3=7892.9 \mathrm{\mu m}$, and arrayed waveguide spacing $d_2= 3 \mathrm{\mu m}$ \cite{tu}. The two AWGs present the same filter shape, and, as a result, the crosstalk in the super-resolution AWG will be higher due to the reduced channel spacing. In the conventional AWG, the theoretical crosstalk is $8$ dB, while the experimental value is $6$ dB \cite{tu}. In the super-resolution AWG, the crosstalk is reduced to approximately $1$ dB, due to the reduction of channel spacing from $5$ GHz to $1$ GHz, without modifying the filter shape. To further enhance crosstalk performance, the development of a new AWG design is necessary and will be addressed in future work. 

The spacing of the output waveguides in the conventional AWG was $d_3= 5 \mathrm{\mu m}$ and the corresponding resolution was $\Delta \lambda=50$ pm. To achieve a resolution of $\Delta \lambda'= 8 $ pm (\textit{i.e.} channel spacing of $1$ GHz), we set $F=19$, $d_3 = 31 \mathrm{\mu m}$, and consequently $d_1=29.37 \mathrm{\mu m}$ (from  Eq. (\ref{d1})). 
A large value of $d_1$ increases the FSR' up to $144$ pm.

The spectra of the light transmitted at the $N=7$ AWG output ports are reported in Fig. \ref{fig4}.

It is important to observe that the super-resolution AWG incorporates an additional Y-branch or slab coupler, which introduces theoretical insertion losses of $10 log_{10} N=-8.45$ dB.

\section{Conclusions}
\label{sec5}
A novel method to enhance the spectral resolution of a custom AWG is presented, and to clarify the underlying principle and the Moir\'e effect, we first review the design concept of a super-resolution spectrometer as proposed by T. Konishi.

Our approach introduces an additional grating at the input plane of a custom AWG. In this configuration, the spectral resolution is determined by the difference in grating periods at the input and output planes, enabling significant improvement without increasing the differential path length. However, the additional splitter or slab coupler increases the device size and introduces additional losses of  $10 log_{10} N$ dB.

We outline the design guidelines for an AWG with $N=7$ ports and $1$ GHz channel spacing, supported by numerical evaluations of the transmission spectra. Future work will focus on optimizing the filter response while addressing apodization effects arising from waveguide mode profiles.

\end{document}